\documentclass[aps,pre,twocolumn,showpacs]{revtex4}

\usepackage{graphicx}

\begin{document}

\title{Instabilities of Shercliffe and Stewartson layers
 in spherical Couette flow}

\author{X. Wei}
\address{Department of Engineering, University of Cambridge,
Cambridge CB2 1PZ, United Kingdom}
\author{R. Hollerbach}
\address{Department of Applied Mathematics, University of Leeds,
Leeds, LS2 9JT, United Kingdom}

\date{\today}

\begin{abstract}
We explore numerically the flow induced in a spherical shell by
differentially rotating the inner and outer spheres.  The fluid
is also taken to be electrically conducting (in the low magnetic
Reynolds number limit), and a magnetic field is imposed parallel to
the axis of rotation.  If the outer sphere is stationary, the magnetic
field induces a Shercliffe layer on the tangent cylinder, the cylinder
just touching the inner sphere and parallel to the field.  If the
magnetic field is absent, but a strong overall rotation is present,
Coriolis effects induce a Stewartson layer on the tangent cylinder.
The non-axisymmetric instabilities of both types of layer separately
have been studied before; here we consider the two cases side by side,
as well as the mixed case, and investigate how magnetic and rotational
effects interact.  We find that if the differential rotation and the
overall rotation are in the same direction, the overall rotation may
have a destabilizing influence, whereas if the differential rotation
and the overall rotation are in the opposite direction, the overall
rotation always has a stabilizing influence.

\end{abstract}


\maketitle

\section{Introduction}

The study of free shear layers and their instabilities is one of
the oldest problems in fluid dynamics, dating back to the pioneering
work of Kelvin \cite{KE} and Helmholtz \cite{HE}.  In this work we
will consider two types of shear layer, the magnetically induced
Shercliffe layer and the rotationally induced Stewartson layer, that
can easily be set up in a differentially rotating spherical shell.
Previous work has studied each of these layers and its corresponding
instabilities in isolation \cite{HS,H1}.  The basic shear layers are
similar in many ways, but their instabilities may be very different.
Here we compare and contrast the two problems, and then consider the
mixed case, when both magnetic and rotational effects are present.

Previous studies on magnetohydrodynamic spherical Couette flow have
included analytic \cite{STA,SO,DO1}, numerical \cite{H2,DO2,HCF} and
experimental \cite{LA,DTS} work, with a variety of imposed magnetic
fields.  Nonmagnetic Stewartson layers have also been widely studied,
in both cylindrical \cite{ST1,HT,BU,FR,AR} and spherical
\cite{ST2,HFME,SC} geometries.

\section{Equations}

We start with two concentric spheres, of radii $r_i$ and $r_o$,
rotating about a common axis (the $z$ axis) with angular velocities
$\Omega_i$ and $\Omega_o$.  The fluid filling the shell is taken to
be electrically conducting, and a magnetic field ${\bf B}=B_0
{\bf\hat e}_z$ is externally imposed.  The question then is, what
sort of flow states will result as the parameters $B_0$, $\Omega_i$
and $\Omega_o$ are varied, and can the solutions be classified in
some systematic way, for example according to whether they are
magnetically or rotationally dominated?

In the reference frame rotating with the outer sphere, the governing
equations are
$$\frac{\partial{\bf U}}{\partial t}+{\rm Re}\,{\bf U\cdot\nabla U}
  + {\rm Ta}\,{\bf\hat e}_z\times{\bf U}$$
$$=-\nabla p + \nabla^2{\bf U}
  + {\rm Ha}^2\,{\bf(\nabla\times b)\times\hat e}_z,\eqno(1)$$
$$\nabla^2{\bf b}=-\nabla\times({\bf U\times\hat e}_z),\eqno(2)$$
where the Hartmann number
$${\rm Ha}=\frac{B_0\,r_i}{\sqrt{\mu\rho\nu\eta}}\eqno(3)$$
measures the strength of the imposed magnetic field, the Taylor number
$${\rm Ta}=\frac{2\,\Omega_o r_i^2}{\nu}\eqno(4)$$
measures the overall rotation of the whole system, and the Reynolds
number
$${\rm Re}=\frac{(\Omega_i-\Omega_o)\,r_i^2}{\nu}\eqno(5)$$
measures the differential rotation of the inner sphere.  The density
$\rho$, viscosity $\nu$, diffusitivy $\eta$, and permeability $\mu$ are
(constant) material properties of the fluid.

In these equations, length has been scaled by $r_i$, time by $r_i^2/\nu$,
and $\bf U$ by $(\Omega_i-\Omega_o)r_i$.  Finally, the induced magnetic
field $\bf b$ has been scaled by ${\rm Rm}\,B_0$, where ${\rm Rm}=
(\Omega_i-\Omega_o)\,r_i^2/\eta$ is the magnetic Reynolds number,
and Eqs.\ (1) and (2) have been formulated in the ${\rm Rm}\to0$ limit,
in which ${\rm Rm}$ no longer appears in the equations at all, but only
in the interpretation associated with $\bf b$.  See also \cite{HS}; the
problem considered here is precisely the extension of \cite{HS} to include
the overall rotation given by ${\rm Ta}$.

The boundary conditions associated with (1) are the usual spherical
Couette flow conditions
$${\bf U}=r\sin\theta\,{\bf\hat e}_\phi\quad{\rm at}\ r=r_i,\quad
 {\bf U}={\bf 0}\quad{\rm at}\ r=r_o,\eqno(6)$$
where the radii will be fixed at $r_i=1$ and $r_o=3$.  For the boundary
conditions associated with (2) we take the exterior regions $r<r_i$
and $r>r_o$ to be insulating.  As shown in \cite{HS}, taking these
regions to be conducting instead can have dramatic consequences,
yielding a counter-rotating jet rather than a shear layer.  However,
precisely because we want to focus on shear layers here, we consider only
the insulating case.

These equations and associated boundary conditions are solved numerically,
using the spherical harmonics code \cite{IJNMF}.  We begin by considering
the axisymmetric basic states, then we linearize about these solutions, and
compute the linear onset of non-axisymmetric instabilities.  Resolutions
as high as 300 Legendre functions in $\theta$ and 180 Chebyshev polynomials
in $r$ were used, and results were tested to ensure that all aspects of the
solutions were fully resolved.

\section{The two pure cases}

\subsection{Basic States}

Figure 1 shows the solutions at ${\rm Re}={\rm Ta}=0$, and ${\rm Ha}^2
=10^4$, $10^5$, $10^6$, corresponding to an infinitesimal differential
rotation, no overall rotation, and an increasingly strong magnetic field.
We see the emergence of an increasingly thin shear layer, the Shercliffe
layer, located on the so-called tangent cylinder $\mathcal C$, the cylinder
just touching the inner sphere and parallel to the magnetic field.  The
origin of this layer is easy to understand, in terms of the magnetic
tension along the field lines.  Fluid columns outside $\mathcal C$ are
coupled at both endpoints to the outer boundary only, so they remain at
rest.  In contrast, fluid columns inside $\mathcal C$ are coupled to both
boundaries, which are rotating at different angular velocities, 0 at the
outer boundary, and 1 at the inner boundary, as imposed by (6).  These
columns then rotate at a rate intermediate between 0 and 1.  The result
is a jump in angular velocity across $\mathcal C$, which is precisely the
Shercliffe layer observed in Figure 1.  (Inside $\mathcal C$ there are
also Hartmann layers at the outer and inner boundaries, accomodating the
jump from $\sim1/2$ in the interior to 0 and 1 at the boundaries.  We
will not be interested in these boundary layers though.)

Figure 2 shows the solutions at ${\rm Re}={\rm Ha}=0$, and ${\rm Ta}
=10^{3.5}$, $10^4$, $10^{4.5}$.  We again see the emergence of an
increasingly thin shear layer, the Stewartson layer, on the same tangent
cylinder $\mathcal C$ as before.  The origin of this layer is also very
similar to that of the Shercliffe layer, the only difference being that
now it is the Taylor-Proudman theorem that couples fluid columns along
the $z$ axis, and not magnetic tension, which is of course entirely absent
for ${\rm Ha}=0$.

Despite their similarities, there are also important differences
between Shercliffe and Stewartson layers.  Note for example how the contour
lines in the Stewartson layer are almost perfectly parallel, whereas in the
Shercliffe layer they spread out somewhat away from the inner sphere.
Related to this is the fact that the asymptotics of these two shear layers
are also slightly different; the Shercliffe layer consists of a single
layer of thickness ${\rm Ha}^{-1/2}$ \cite{RO}, whereas the Stewartson layer
consists of a primary layer of thickness ${\rm Ta}^{-1/4}$ across which the
shear is resolved, but also contains secondary layers of thicknesses
${\rm Ta}^{-2/7}$ just inside $\mathcal C$ and ${\rm Ta}^{-1/3}$ just
outside $\mathcal C$ \cite{ST2}.

\begin{figure}
\includegraphics[scale=0.95]{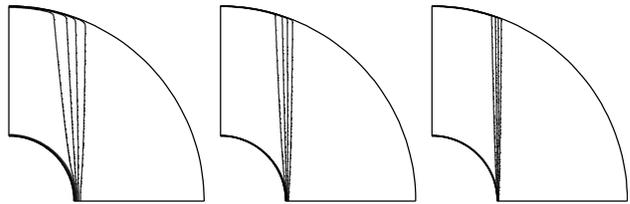}
\caption{Examples of the pure Shercliffe layer, showing
contours of the angular velocity, with a contour interval of 1/9.
From left to right ${\rm Ha}^2=10^4$, $10^5$, $10^6$, and ${\rm Ta}=
{\rm Re}=0$ for all three.}
\end{figure}

\begin{figure}
\includegraphics[scale=0.95]{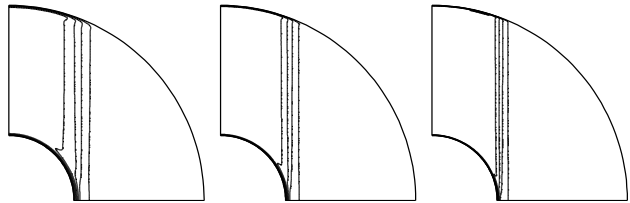}
\caption{Examples of the pure Stewartson layer, showing
contours of the angular velocity, with a contour interval of 1/9.
From left to right ${\rm Ta}=10^{3.5}$, $10^4$, $10^{4.5}$, and
${\rm Ha}={\rm Re}=0$ for all three.}
\end{figure}

\subsection{Onset of Instabilities}

The results in Figs.\ 1 and 2 are all for the case of infinitesimally
small differential rotation, ${\rm Re}=0$.  Now suppose the differential
rotation is gradually increased.  As the shear across the layers is
increased, one might expect the layers to become unstable eventually, to
something like a Kelvin-Helmholtz instability.  That is, one might expect
the initially circular, axisymmetric basic state to adopt a wavy,
non-axisymmetric structure.

The instabilities of the pure Shercliffe layer were considered by \cite{HS};
the left panel in Fig.\ 3 shows their results (Fig.\ 4a in \cite{HS}), over
the range of Hartmann numbers shown in Fig.\ 1.  The instabilities of the
pure Stewartson layer were considered by \cite{H1}; the right panel in
Fig.\ 3 shows these results (Fig.\ 4 in \cite{H1}), again over the range of
Taylor numbers shown in Fig.\ 2.  (Note though that the equations here are
scaled differently from those in \cite{H1}, to allow for the possibility of
${\rm Ta}=0$ here.  The Ekman and Rossby numbers in \cite{H1} are related
to the Taylor and Reynolds numbers used here by ${\rm E}=1/(2{\rm Ta})$
and ${\rm Ro}=2{\rm Re}/{\rm Ta}$.)

\begin{figure}
\includegraphics[scale=0.90]{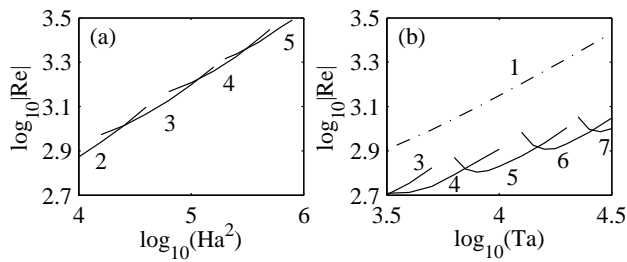}
\caption{(a) shows $\log_{10}|{\rm Re_c}|$ as a function of
$\log_{10}({\rm Ha})$ for the pure Shercliffe layer; (b) shows
$\log_{10}|{\rm Re_c}|$ as a function of $\log_{10}({\rm Ta})$ for the
pure Stewartson layer.  The numbers beside individual curves indicate the
azimuthal wavenumbers $m$, showing only the most unstable modes.  In (b),
the solid curves $m=3$ to $7$ are for ${\rm Re}>0$, whereas the dashed
curve $m=1$ is for ${\rm Re}<0$.}
\end{figure}

Comparing the two panels in Fig.\ 3, there are clearly similarities
between the two cases.  Most obviously, for comparable thicknesses of
the underlying shear layers (as indicated in Figs.\ 1 and 2) the critical
Reynolds numbers for the onset of instabilities are also comparable, with
$|{\rm Re_c}|\sim10^3$ for the ${\rm Ha}$ and ${\rm Ta}$ ranges shown.
Beyond that, in both cases the critical Reynolds numbers increase as the
thickness of the shear layers decreases.

However, there is also one crucial difference between the instabilities
of these two types of shear layer, more fundamental than any difference
between the shear layers themselves.  Specifically, for the pure
Shercliffe layer, the results are invariant to the sign of ${\rm Re}$,
that is, the direction in which the inner sphere rotates.  The easiest
way to see this is to note that reversing the rotation of the inner sphere
is equivalent to turning the entire system upside down.  This merely
reverses the sign of the imposed magnetic field though, which clearly has
no effect.

In sharp contrast, the results for the Stewartson layer are not invariant
to the sign of ${\rm Re}$.  One can of course still imagine turning
the system upside down, but instead of reversing the magnetic field, this
now reverses the sense of the overall rotation.  And unlike the magnetic
field, whose sign doesn't matter, the sign of the overall rotation does
matter.  That is, instead of reversing the sign of ${\rm Re}$ one could just
as well reverse the sign of ${\rm Ta}$, but the result is still not
equivalent to the original configuration.  Having ${\rm Re}$ and ${\rm Ta}$
of the same sign is fundamentally different from having them of the opposite
sign.

Returning to the right panel in Fig.\ 3 then, we note that ${\rm Re}>0$ and
${\rm Re}<0$ do indeed yield strikingly different instabilities.  Positive
${\rm Re}$ has increasingly large azimuthal wavenumbers $m$ for increasingly
large ${\rm Ta}$, exactly as one would expect for a Kelvin-Helmholtz type
instability, whereas negative ${\rm Re}$ has $m=1$ over the entire range of
${\rm Ta}$ shown here.  Comparing with the $\pm {\rm Re}$ invariant
Shercliffe results in the left panel, we see that these are more like the
${\rm Re}>0$ Stewartson results, in that they also show a progression to
higher $m$.

One reason for considering the mixed Shercliffe-Stewartson problem then
is simply to see how this $\pm {\rm Re}$ asymmetry manifests itself in this
case, and at what point the results are more like the symmetric Shercliffe
problem or more like the asymmetric Stewartson problem.

\subsection{Location of the Instabilities}

\begin{figure}
\includegraphics[scale=0.95]{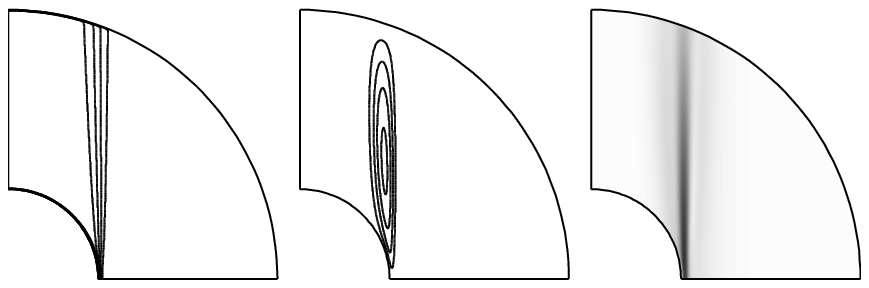}
\caption{The Shercliffe layer at ${\rm Ha}^2=10^5$, ${\rm Re_c}=\pm1588$,
and ${\rm Ta}=0$.  The left panel shows the angular velocity, with a
contour interval of 1/9; the middle panel shows the meridional circulation,
with a contour interval of $10^{-4}$; the right panel shows the azimuthally
integrated kinetic energy of the instability, having wavenumber $m=3$, as
indicated in the left panel of Fig.\ 3.}
\end{figure}

\begin{figure}
\includegraphics[scale=0.95]{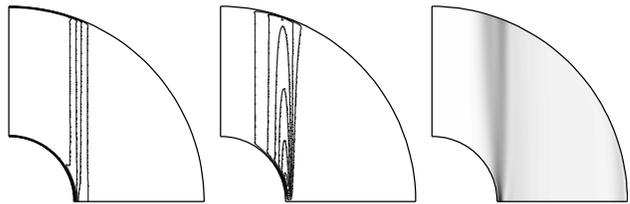}
\caption{The Stewartson layer at ${\rm Ta}=10^4$, ${\rm Re_c}=665$,
and ${\rm Ha}=0$.  The left panel shows the angular velocity, with a
contour interval of 1/9; the middle panel shows the meridional circulation,
with a contour interval of $10^{-3}$; the right panel shows the azimuthally
integrated kinetic energy of the instability, having wavenumber $m=5$, as
indicated in the right panel of Fig.\ 3.}
\end{figure}

First though we consider a few more aspects of the two pure problems,
namely the spatial location of the instabilities.  Fig.\ 4 shows the
Shercliffe case, Fig.\ 5 the ${\rm Re}>0$ Stewartson case.  The first panels
in each figure show the angular velocity, as before in Figs.\ 1 and 2.  One
point to note here is how similar these solutions at non-zero ${\rm Re}$ are
to the ${\rm Re}=0$ solutions in Figs.\ 1 and 2.  The inertial term
${\rm Re} \bf U\cdot\nabla U$ is crucially important in driving the
instabilities, but in the basic states themselves it is almost
completely balanced by the pressure-gradient term.  The second panels
show the associated meridional circulation.  This is very weak though in
comparison with the shear layers, and does not appear to play an important
role in the instabilities.  Finally, the grey-shading in the third panels
shows the azimuthally integrated kinetic energy of the instabilities, that
is, the quantity $\int|{\bf u}|^2\,r\sin\theta\,d\phi$.  As expected, both
instabilities are concentrated on the tangent cylinder $\mathcal C$,
although it is interesting to note that the concentration is far greater
in the Shercliffe case than in the Stewartson case.

Figure 6 shows the corresponding results for the ${\rm Re}<0$ Stewartson
case, the anomalous $m=1$ mode.  The instability now appears to have
curious gaps in cylindrical radius, resulting in a striped appearance.
Furthermore, the instability reaches its maximum concentration not on
$\mathcal C$, but instead just inside, where the Stewartson layer
intersects the Ekman layer on the inner sphere.  This would suggest
that this anomalous mode is perhaps not a Stewartson layer instability
at all, but instead an Ekman layer instability, for which it is well
known that the ${\rm Re}>0$ case (von K\'arm\'an flow) and the
${\rm Re}<0$ case (B\"odewadt flow) are indeed very different \cite{JG}.

\begin{figure}
\includegraphics[scale=0.95]{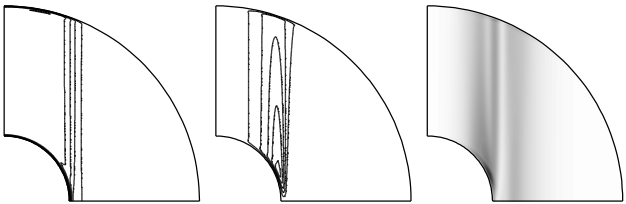}
\caption{As in Fig.\ 5, but with ${\rm Re_c}=-1404$, and $m=1$.}
\end{figure}

\begin{figure}
\includegraphics[scale=0.95]{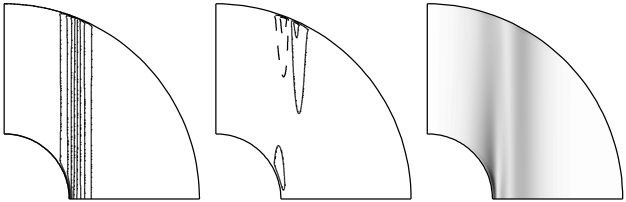}
\caption{As in Fig.\ 6, but with the split outer sphere boundary
condition.  ${\rm Re_c}=-826$, and $m=1$.}
\end{figure}

The numerical tests conducted by \cite{H1} considered exactly this
possibility, and suggest that however plausible this idea may be, it is
incorrect: this ${\rm Re}<0$ mode is not a B\"odewadt instability, but a
Stewartson layer instability, just like the ${\rm Re}>0$ modes.  However,
these numerical tests \cite{H1}, in which the meridional circulation
and the Ekman layers were simply deleted from the basic state before
computing the instabilities, can be -- and indeed have been --
criticized as being unphysical, not corresponding to anything that
one could actually set up in a lab, for example.

We would therefore like to repeat something like this deletion of the
meridional circulation and the Ekman layers, but in a way that is
physical, and could be set up in an experiment.  Fortunately, this can
be accomplished rather easily: we simply replace the outer boundary
condition $\bf U=0$ by ${\bf U}=r\sin\theta\,\Omega(\theta)\,{\bf\hat e}
_\phi$ at $r=r_o$, where $\Omega(\theta)$ is 1 inside $\mathcal C$,
and 0 outside (for numerical reasons the transition is actually
smoothed out over a degree or so).  Physically this would correspond
to having the outer sphere split into differentially rotating segments,
with the regions inside the tangent cylinder now co-rotating with the
inner sphere, which is precisely how many Stewartson layer experiments
are indeed done \cite{FR,SC}.

Figure 7 shows these results.  If we begin by comparing the basic
states in Figs.\ 6 and 7, we see that this new boundary condition
has roughly doubled the jump in angular velocity across the shear,
because everything inside $\mathcal C$ is now co-rotating with the
inner sphere, whereas before the fluid inside $\mathcal C$ was rotating
at a rate intermediate between 1 at the inner boundary and 0 at the
outer.  Turning next to the Ekman layers, these have been largely
eliminated; if everything inside $\mathcal C$ is co-rotating, there is
simply no need for Ekman layers at the boundaries.  And correspondingly,
the meridional circulation, which is driven by Ekman pumping in the
boundary layers, is also dramatically reduced.

This new boundary condition has thus accomplished exactly what we
wanted, but in a way that is physically realizable, unlike the numerical
tests presented in \cite{H1}.  And if we compare the instabilities in
the two cases then, the original boundary condition (Fig.\ 6) has $m=1$
and ${\rm Re_c}=-1404$, and the new boundary condition (Fig.\ 7) has $m=1$
and ${\rm Re_c}=-826$.  That is, doubling the shear across the layer roughly
halves the critical Reynolds number, exactly as one would expect if it is
indeed the shear layer that is triggering the instability.  This new
boundary condition therefore confirms the claim made by \cite{H1} that this
anomalous $m=1$ mode is also a Stewartson layer instability (although
beyond that there are unfortunately still many aspects of this mode that
are not entirely clear).

\section{The mixed case}

Figure 8 shows stability results for the mixed case, when neither
${\rm Ha}$ nor ${\rm Ta}$ are zero.  Starting with Hartmann numbers
${\rm Ha}^2=10^{4.5}$ and $10^{5.5}$, ${\rm Ta}$ is increased from $10^2$
to $10^{5.5}$, for both positive and negative ${\rm Re}$.  For ${\rm Ta}
\ll {\rm Ha}^2$, $\pm {\rm Re}$ are not surprisingly almost the same.  As
rotational effects become comparable with magnetic effects though, the
rotationally induced asymmetry becomes more and more pronounced.  For
positive/negative ${\rm Re}$ the wavenumber increases/decreases, until for
${\rm Ta}=10^{5.5}$ we are almost back to the pure Stewartson regime, with
large $m$ for positive ${\rm Re}$, and small $m$ for negative ${\rm Re}$.
It is unfortunately not entirely clear why the wavenumbers behave in this
way, but the fact that there is this smooth transition from the symmetric
Shercliffe case to the asymmetric Stewartson cases, for both positive and
negative ${\rm Re}$, further reinforces the view that even these
`anomalous' $m=1$ modes above are not so anomalous after all, but are
merely the limiting case in this family of shear layer instabilities.

One other interesting, and completely unexpected result in Fig.\ 8 is
this initial decrease in the ${\rm Re}>0$ curves, reaching a minimum when
${\rm Ta}/{\rm Ha}^2\sim10^{-0.75}\approx0.2$.  ${\rm Ta}/{\rm Ha}^2=O(1)$
is of course precisely the regime where rotational and magnetic effects are
comparable, so it is not surprising that any interaction between the two
would manifest itself most strongly there.  That rotational and magnetic
effects can destabilize one another even though each separately has a
stabilizing influence is also familiar in other contexts, such as
Rayleigh-B\'enard convection \cite{CH}.  It is not clear though why this
mutual destabilization in this case does not occur for ${\rm Re}<0$ as well.

\begin{figure}
\includegraphics[scale=0.9]{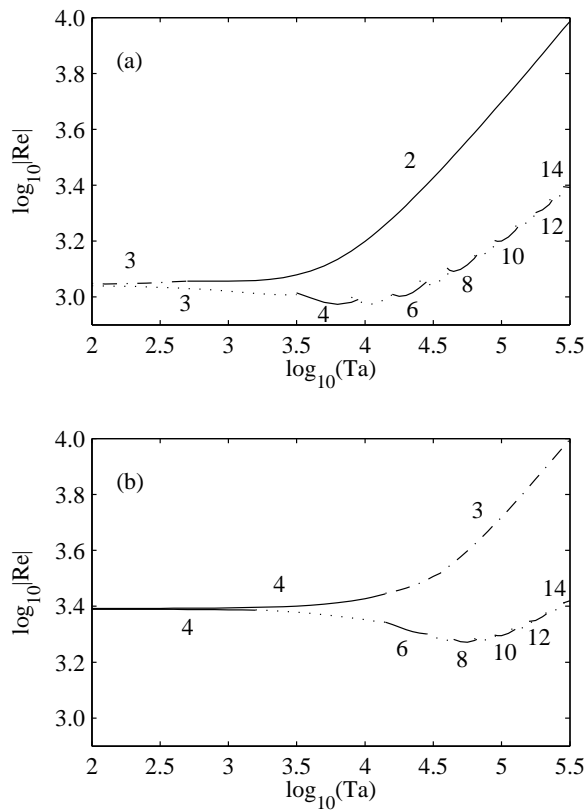}
\caption{The critical Reynolds numbers for the onset of instabilities,
as functions of $\log_{10}({\rm Ta})$, with ${\rm Ta}=10^2$ almost the
pure Shercliffe regime, and ${\rm Ta}=10^{5.5}$ almost the pure Stewartson
regime.  (a) is for ${\rm Ha}^2=10^{4.5}$; (b) is for ${\rm Ha}^2=10^{5.5}$.
Within each panel the upper set of curves, with decreasing wavenumbers, is
for ${\rm Re}<0$; the lower set of curves, with increasing wavenumbers, is
for ${\rm Re}>0$.}
\end{figure}

Finally, we wish to consider the spatial structures of both the basic
states and the instabilities in this case ${\rm Ta}/{\rm Ha}^2=10^{-0.75}$
where rotational and magnetic effects are interacting most strongly, and
see whether they are more like the pure Shercliffe case, or more like the
pure Stewartson case.  Figure 9 shows the results for ${\rm Re}>0$;
comparing with Figs.\ 4 and 5, we see that they look more like the pure
Shercliffe case.  Figure 10 shows the results for ${\rm Re}<0$; comparing
with Figs.\ 4 and 6, we see that the basic state again looks more like the
pure Shercliffe case.  The instabilities though have aspects in common with
both the Stewartson case, namely this striped appearance, as well as the
Shercliffe case, namely the concentration more outside $\mathcal C$,
rather than in the equatorial region as in the Stewartson case.

\begin{figure}
\includegraphics[scale=0.95]{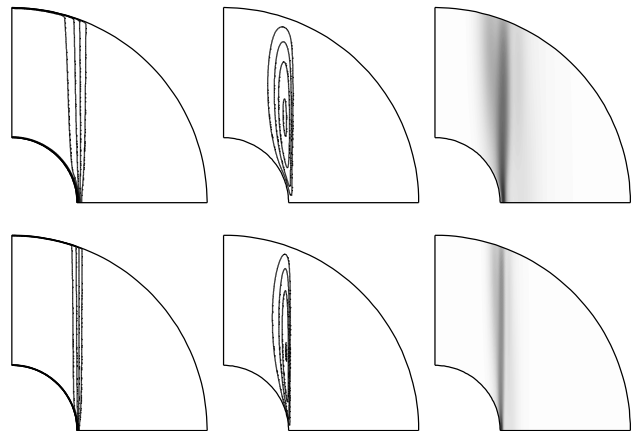}
\caption{The top row shows results at ${\rm Ha}^2=10^{4.5}$, ${\rm Ta}
=10^{3.75}$, ${\rm Re_c}=946$ and $m=4$, the bottom row at ${\rm Ha}^2
=10^{5.5}$, ${\rm Ta}=10^{4.75}$, ${\rm Re_c}=1827$ and $m=8$; that is, at
the minima of the ${\rm Re}>0$ curves in Fig.\ 8.  The left panels show the
angular velocity, with a contour interval of 1/9; the middle panels show
the meridional circulation, with a contour interval of $10^{-3}$ (top) and
$5\times10^{-4}$ (bottom); the right panels show the azimuthally integrated
kinetic energy of the instabilities.}
\end{figure}

\begin{figure}
\includegraphics[scale=0.95]{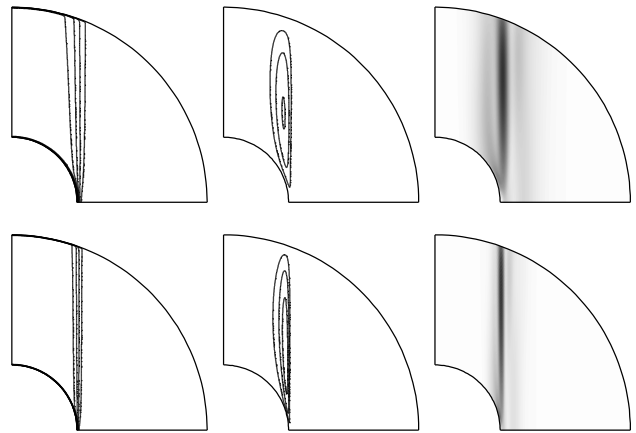}
\caption{As in Fig.\ 9, but for ${\rm Re}<0$, with ${\rm Ha}^2=10^{4.5}$,
${\rm Ta}=10^{3.75}$, ${\rm Re_c}=-1324$ and $m=2$ for the top row, and
${\rm Ha}^2=10^{5.5}$, ${\rm Ta}=10^{4.75}$, ${\rm Re_c}=-3661$ and $m=3$
for the bottom row.}
\end{figure}

\section{Conclusion}

In this work we have explored the stability of two types of free shear
layers that may be set up by magnetic and rotational effects.  Although
the shear layers themselves are very similar for the two effects, the
instabilities are quite different in one important aspect, namely that
in the magnetic Shercliffe case they are invariant to the sign of the
differential rotation that induces them, whereas in the rotational
Stewartson case they are not.  However, as different as the $\pm {\rm Re}$
pure Stewartson cases may at first sight appear to be, by considering
the mixed case, we showed that there is in fact a smooth progression
from the invariant Shercliffe limit to both of the $\pm {\rm Re}$
Stewartson cases, suggesting that these cases are not so different
after all.


\begin{thebibliography}{}
\bibitem{KE}
Lord Kelvin, Phil.\ Mag.\ {\bf42}, 362 (1871).
\bibitem{HE}
H. von Helmholtz, Phil.\ Mag.\ {\bf36}, 337 (1868).
\bibitem{HS}
R. Hollerbach and S. Skinner,
Proc.\ R. Soc.\ Lond.\ A {\bf457}, 785 (2001).
\bibitem{H1}
R. Hollerbach, J. Fluid Mech.\ {\bf492}, 289 (2003).
\bibitem{STA}
S.V. Starchenko, Phys.\ Fluids {\bf10}, 2414 (1998).
\bibitem{SO}
N. Kleeorin, I. Rogachevskii, A. Ruzmaikin, A. Soward
and S. Starchenko, J. Fluid Mech.\ {\bf344}, 213 (1997).
\bibitem{DO1}
E. Dormy, D. Jault and A.M. Soward, J. Fluid Mech.\ {\bf452},
263 (2002).
\bibitem{H2}
R. Hollerbach, Proc.\ R. Soc.\ Lond.\ A {\bf444}, 333 (1994).
\bibitem{DO2}
E. Dormy, P. Cardin and D. Jault,
Earth Planet Sci.\ Lett.\ {\bf160}, 15 (1998).
\bibitem{HCF}
R. Hollerbach, E. Canet and A. Fournier,
Europ.\ J. Mech.\ B {\bf26}, 729 (2007).
\bibitem{LA}
D.R. Sisan, N. Mujica, W.A. Tillotson, Y.M. Huang, W. Dorland,
A.B. Hassam, T.M. Antonsen, D.P. Lathrop, Phys.\ Rev.\ Lett.\ {\bf93},
114502 (2004).
\bibitem{DTS}
H.-C. Nataf, T. Alboussiere, D. Brito, P. Cardin, N. Gagniere,
D. Jault, J.-P. Masson and D. Schmitt, Geophys.\ Astrophys.\ Fluid
Dynam.\ {\bf100}, 281 (2006).
\bibitem{ST1}
K. Stewartson, J. Fluid Mech.\ {\bf3}, 17 (1957).
\bibitem{HT}
R. Hide and C.W. Titman, J. Fluid Mech.\ {\bf29}, 39 (1967).
\bibitem{BU}
F.H. Busse, J. Fluid Mech.\ {\bf33}, 577 (1968).
\bibitem{FR}
W.G. Fr\"uh and P.L. Read, J. Fluid Mech.\ {\bf383}, 143 (1999).
\bibitem{AR}
A. Aguiar and P. Read, Met.\ Z. {\bf15}, 417 (2006).
\bibitem{ST2}
K. Stewartson, J. Fluid Mech.\ {\bf26}, 131 (1966).
\bibitem{HFME}
R. Hollerbach, B. Futterer, T. More and C. Egbers,
Theor.\ Comp.\ Fluid Dynam.\ {\bf18}, 197 (2004).
\bibitem{SC}
N. Schaeffer and P. Cardin, Phys.\ Fluids {\bf17}, 104111 (2005).
\bibitem{IJNMF}
R. Hollerbach, Int.\ J. Numer.\ Meth.\ Fluids {\bf32}, 773 (2000).
\bibitem{RO}
P.H. Roberts, Proc.\ R. Soc.\ Lond.\ A {\bf300}, 94 (1967).
\bibitem{JG}
H.A. Jasmine and J.S.B. Gajjar, Phil.\ Trans.\ R. Soc.\ Lond.\ A
{\bf363}, 1131 (2005).
\bibitem{CH}
S. Chandrasekhar, {\it Hydrodynamic and Hydromagnetic Stability},
Clarendon, Oxford (1961).
\end{thebibliography}
\end{document}